\documentstyle[prl,multicol,aps,epsfig]{revtex}
\newcommand \beq{\begin{eqnarray}}
\newcommand \eeq{\end{eqnarray}}
\newcommand \bea{\begin{eqnarray}}
\newcommand \eea{\end{eqnarray}}
\newcommand \ga{\raisebox{-.5ex}{$\stackrel{>}{\sim}$}}
\newcommand \la{\raisebox{-.5ex}{$\stackrel{<}{\sim}$}}

\def\simge{\mathrel{%
       \rlap{\raise 0.511ex \hbox{$>$}}{\lower 0.511ex \hbox{$\sim$}}}}
\def\simle{\mathrel{
       \rlap{\raise 0.511ex \hbox{$<$}}{\lower 0.511ex \hbox{$\sim$}}}}

\begin{document}

    \title{Vortex core structure and global properties of rapidly rotating
Bose-Einstein condensates}
\author{Gordon Baym and C.J. Pethick}
\address{Department of Physics, University of Illinois at
Urbana-Champaign, 1110 West Green Street, Urbana, Illinois 61801, \\
and\\
NORDITA, Blegdamsvej 17, DK-2100 Copenhagen \O, Denmark}

\maketitle

\begin{abstract}

    We develop an approach for calculating stationary states of rotating
Bose-Einstein condensates in harmonic traps which is applicable for arbitrary
ratios of the rotation frequency to the transverse frequency of the trap
$\omega_{\perp}$.  Assuming the number of vortices to be large, we write the
condensate wave function as the product of a function that describes the
structure of individual vortices times an envelope function, varying slowly on
the scale of the vortex spacing.  By minimizing the energy, we derive
Gross-Pitaevskii equations that determine the properties of individual
vortices and the global structure of the cloud.  For low rotation rates, the
structure of a vortex is that of an isolated vortex in a uniform medium, while
for rotation rates approaching the frequency of the trap (the mean field
quantum Hall regime), the structure is that of the lowest p-wave state of a
particle in a harmonic trap with frequency $\omega_{\perp}$.  The global
structure of the cloud is determined by minimizing the energy with respect to
variations of the envelope function; for conditions appropriate to most
experimental investigations to date, we predict that the transverse density
profile of the cloud will be of the Thomas-Fermi form, rather than the
Gaussian structure predicted on the assumption that the wave function consists
only of components in the lowest Landau level.

\end{abstract}

\section{Introduction}

    Bose-Einstein-condensed atomic gases are very well suited to investigating
quantized vortex lines.  Single vortex lines were first made in atomic
condensates by Matthews et al.  \cite{JILA}, who induced rotation by phase
imprinting in a spinor condensate.  Subsequently, arrays containing many
vortices were created in scalar condensates by inducing rotation mechanically,
either by stirring the condensate \cite{Madison,VortexLatticeBEC}, or by
evaporating particles \cite{HaljanCornell}.  For a theoretical review, see
Ref.\cite{fetterreview}.  In a seminal work, Ho \cite{Ho} predicted that
clouds of atoms confined in harmonic traps, when rotated at frequencies close
to the transverse frequency $\omega_{\perp}$ of the trap, should condense into
the lowest Landau level (LLL) in the Coriolis force, similar to charged
particles in the quantum Hall regime.  This insight has led to extensive
experimental studies in which rotation rates in excess of 0.99
$\omega_{\perp}$ have been achieved, and the structure of the condensate
within a single cell of the vortex lattice has been examined
\cite{HaljanCornell,jila2}.

    To date, most theoretical work on vortices in harmonically trapped
condensates rotating at frequencies close to $\omega_{\perp}$ has been based
on the use of wave functions in which particles occupy only the lowest Landau
level.  In contrast, for slowly rotating condensates, the usual approach to
calculating vortex structure is to solve the Gross-Pitaevskii equation.  In
this paper we address the question of how this approach goes over to the mean
field quantum Hall description when the rotation rate is increased
\cite{validity}.  We develop a unified method for calculating both the
structure of individual vortices and the global structure of the cloud for
arbitrary rotation rates.  Writing the condensate wave function as a product
of a slowly varying envelope function that determines the density averaged
over a single cell of the vortex lattice, and a function that determines the
variations of the wave function on length scales of order the vortex
separation and core size, we derive, in Sec.~II, the energy of the system.
Then in Sec.~III we derive the equation for the structure of the wave function
within a single cell of the vortex lattice by variation of the energy
functional.  In Sec.~IV we derive equations for the global structure of the
cloud.  We find that a wave function made up only of components belonging to
the lowest Landau level is a good approximation only under a very restricted
range of conditions; if in the non-rotating system the density profile in the
plane transverse to the rotation axis is of the Thomas-Fermi form, an inverted
parabola, then at high rotation the shape remains Thomas-Fermi, and thus
includes components from higher Landau levels, rather than the Gaussian shape
predicted by the LLL calculation.

\section{Basic Formalism}

    We consider a system of weakly interacting bosons trapped in a harmonic
potential, $V(r) = \frac12 m (\omega^2 r_\perp^2+\omega_z z^2)$, where $\vec
r_\perp=(x,y)$, rotating at angular velocity $\Omega$ about the $z$ axis.  The
angular momentum of the system is carried in quantized vortices, of number
$N_v \gg 1$ at large rotation rates.  We assume the vortices to be rectilinear
and to form a triangular lattice.  When $N_v\gg 1$ the rotation rate of the
lattice is related to the (two dimensional) density of vortices, $n_v$, by
\beq
  m\Omega = \pi \hbar n_v,
\label{omeganv}
\eeq
as follows from identifying the mean velocity via the condition
for quantization of vorticity, $\oint_{\cal C} d{\vec \ell}\cdot {\vec v}=
hN_v({\cal C})/m$, where $N_v({\cal C})$ is the number of vortices surrounded
by the contour $\cal C$.  We will not address here the form of the corrections
to Eq.~(\ref{omeganv}) for finite $N_v$ \cite{corrections}.

    In order to separate out the short distance vortex structure from the
large scale structure, we follow the approach of Fischer and Baym \cite{FG}
and write the order parameter as
\beq
  \psi(\vec r\,) =  e^{i\Phi(\vec r\,)}f(\vec r\,) \sqrt{n(\vec r\,)},
 \label{psi}
\eeq
the product of a rapidly varying real factor, $f(\vec r\,)$, which
vanishes at each vortex core, times a slowly varying real envelope function,
$\sqrt{n(\vec r\,)}$, and a phase factor.  We normalize $f^2$ so that it
averages to unity over each unit cell of the lattice; thus $n(\vec r\,)$ is
the smoothed density profile of the system, which varies slowly over the unit
cells of the vortex lattice.  The factor $e^{i\Phi}f$ describes the local
swirling of the fluid -- with the phase $\Phi$ wrapping by 2$\pi$ around each
vortex -- together with the overall rotation of the vortex lattice at
$\Omega$.  We generally set $\hbar=1$.

    The total energy of the system in the laboratory frame is
\beq
 E=  \int d^3r \left\{\frac{\hbar^2}{2m}|\nabla\psi|^2 +
   V(r)n(r)f^2(r)+ \frac g2 n^2(r)f^4(r)\right\},
 \label{energy}
\eeq
where we assume a two body interaction described by an s-wave scattering
length, $a_s$, with $g=4\pi a_s\hbar^2/m$.  With Eq.~(\ref{psi}), the kinetic
energy in the laboratory frame becomes
\beq
\int d^3r \frac{1}{2m}|\nabla\psi|^2 \equiv K
  = \int d^3r
\frac{1}{2m}\left\{(\nabla\sqrt{n})^2+(\nabla\Phi)^2nf^2+n(\nabla f)^2
  +\frac12 \nabla f^2\cdot \nabla n)\right\}.
\eeq
We integrate the final term by parts to give $-\frac12\int f^2 \nabla^2
n$; since $n$ varies slowly across a unit cell of the vortex lattice, we may
replace the $f^2$ here by its average (=1) in the cell, so that the integral
gives only a vanishing surface term.  Thus
\beq
  K= \int d^3r
\frac{1}{2m}\left\{(\nabla\sqrt{n})^2+(\nabla\Phi)^2nf^2 + n(\nabla f)^2
  \right\}.
  \label{ke}
\eeq
In the unit cell centered on vortex $j$ at position $\vec R_j$ in the
plane transverse to the rotation axis, the velocity $\nabla\Phi/m$ is the sum
of the solid body rotation of the position of the vortex, $\vec\Omega \times
\vec R_j$, plus the local velocity around the vortex, which we write as
$\nabla \phi_j/m$:
\beq
  \nabla\Phi(r) \simeq m \vec\Omega \times \vec R_j + \nabla \phi_j.
 \label{phase}
\eeq
The $(\nabla\Phi)^2$ term thus becomes
\beq
  \int d^3r \frac{nf^2}{2m}(\nabla\Phi)^2 =
  \sum_j \int_j d^3r nf^2\left\{\frac{(\nabla\phi_j)^2 }{2m}
  + \frac 12 m\Omega^2R_j^2\right\};
  \label{phi2}
\eeq
the integration is over unit cell $j$, and the sum is over all cells.  The
cross term vanishes since in the limit $N_v\gg1$ the average velocity in the
cell measured with respect to the center of the cell vanishes.  In the
Wigner-Seitz approximation, which we employ below, $\phi_j$ becomes the
azimuthal angle measured with respect to the point $\vec R_j$.  Writing within
cell $j$, $\vec R_j = \vec r_\perp - \vec \rho$, the final term in
Eq.~(\ref{phi2}) becomes $\frac12 I \Omega^2 - \sum_j \int_j \frac12 m
\Omega^2 \rho^2 n f^2$, where $I = \int d^3 r mnf^2 r_\perp^2$ is the total
moment of inertia of the system.  Similarly the transverse trapping potential
term becomes $\frac12 I\omega^2$.

    To determine the equilibrium structure, we work in the frame rotating at
angular velocity $\Omega$.  [This procedure is equivalent to determining the
equilibrium structure at fixed angular momentum, $L$, by minimizing the total
energy taking the constraint of fixed $L$ into account by a Lagrange
multiplier, $\Omega$.]  The total angular momentum along the $z$ axis is given
by
\beq L = \int d^3r n(r)f^2(r)\,\left(\vec r \times \nabla\Phi(r)\right)_z.
\eeq
Using Eq.~(\ref{phase}), and again writing in cell $j$, $\vec r_\perp =
\vec R_j+\vec \rho$, we have,
\beq
 L= I\Omega +\sum_j\int d^3 r
  \,nf^2\left\{(\vec \rho \times \nabla\phi_j)_z - m\Omega \rho^2 \right\},
 \label{lcell}
\eeq
since the average position in the unit cell $j$ is $R_j$, and the average
velocity in the local frame of the vortex $j$ vanishes.  The first term is the
angular momentum of the center of mass of the cell, and the second the
intrinsic angular momentum within the cell.  Assembling the pieces,
Eqs.~(\ref{energy}), (\ref{ke}), (\ref{phi2}), and (\ref{lcell}), we have,
\beq
 E' = E-\Omega L &=& \int d^3r
  \left(\frac{(\nabla\sqrt{n})^2}{2m} + \frac{m}{2}n(r)\omega_z^2 z^2 \right)
  +\frac 12(\omega^2 -\Omega^2)I
  \nonumber\\
   && +\sum_j \int_j
  d^3 r n \left[ \frac{(\nabla f)^2}{2m}+ \frac{f^2}{2m} (\nabla\phi_j)^2
  +\frac{m\Omega^2}{2}\rho^2 f^2 n -\Omega(\vec \rho \times
     \nabla\phi_j)_z
  +\frac g2 nf^4\right].
\label{eprime}
\eeq
Expressing $I$ in terms of the moment of inertia, $\bar I
= \int d^3r\, mnr_\perp^2$, of the smoothed density distribution we write
\beq
 E' &=& \int d^3r \left(\frac{(\nabla\sqrt{n})^2}{2m}
       + \frac{m}{2}n(r)\omega_z^2 z^2 \right)
   +\frac 12(\omega_{\perp}^2-\Omega^2)\bar I + \sum_j E_j,
\label{eprime11}
\eeq
where
\beq
 E_j = \int_j d^3 r n \left\{ \frac{(\nabla f)^2}{2m}+ \frac{f^2}{2m}
 (\nabla\phi_j)^2 + \frac{m\omega_{\perp}^2}{2}\rho^2 (f^2 -1)
 +\frac12 m  \Omega^2\rho^2 -\Omega(\vec \rho \times
 \nabla\phi_j)_z +\frac g2 nf^4\right\}
 \label{ecell}
\eeq
is the internal energy within cell $j$.

\section{Equilibrium structure of vortices}

    We turn now to determining the structure of the vortices within the unit
cells.  To do this we introduce the Wigner-Seitz approximation to evaluate the
vortex sum, replacing the hexagonal unit cell by a circle of radius $\ell=
1/(m\Omega)^{1/2}$.  Then $f$ is cylindrically symmetric within each cell.  In
the following, we assume that the vortex spacing is small compared with the
characteristic length scale in the axial direction.  The term in Eq.\
(\ref{ecell}) containing $\partial f/\partial z$ can then be neglected, and
$f$ depends only on the transverse coordinate and the average local density.
In cell $j$, $\phi_j$ becomes the azimuthal angle with respect to the center
of the cell.  Again we write within cell $j$, $\vec r_\perp =\vec R_j +
\vec\rho$, so that $(\nabla\phi_j)^2 = 1/\rho^2$.  Furthermore, $(\vec \rho
\times \nabla \phi_j)_z$ becomes just $\hbar$, so that the angular momentum in
the Wigner-Seitz approximation is,
\beq
  L =  I \Omega + \sum_j \int_j d^3 r nf^2(1-m\Omega\rho^2)=  I\Omega
   + N\left( 1-\frac{\langle \rho^2\rangle}{\ell^2} \right),
\eeq
where $\langle \rho^2\rangle \simeq \ell^2/2$ is the average of $\rho^2$
within a given cell.  We have neglected gradients of the smoothed density
here.  For an incompressible fluid, $ \langle \rho^2\rangle/\ell^2 =1/2$, and
therefore the additional angular momentum per particle in the Wigner-Seitz
approximation is $\hbar/2$, which is close to Tkachenko's result
\cite{Tkachenko,BC} for a triangular lattice in an incompressible fluid,
$(\pi/4\sqrt3)\hbar \simeq 0.453\hbar$.

    In the Wigner-Seitz approximation, $E_j$ becomes,
\beq
  E_j =  \int_j d^3r n\left\{
 \frac{1}{2m}\left[\left(\frac{\partial f}{\partial\rho}\right)^2
  + \frac{f^2}{\rho^2}\right]+
   \frac m2\omega_{\perp}^2\rho^2(f^2-1)  +\frac12 m\Omega^2\rho^2 -\Omega
   + \frac g2 nf^4\right\}.
\eeq
The form of $f$ within each cell is determined by minimizing $E_j$ with
respect to $f$, subject to $\int_j d^2\rho f^2 = 1$, with the boundary
conditions that $f(0)=0$ and $\partial f/\partial \rho = 0$ at $\rho = \ell$.
Since there are no terms coupling $f$ at different values of $z$, the
equilibrium $f$ depends on $z$ only through the dependence of the average
density on $z$.  Thus within a given cell, at given height $z$,
\beq
  \frac{1}{2m}\left(-\frac1\rho \frac{\partial}{\partial \rho}\left(\rho \frac
 {\partial f}{\partial\rho}\right) +\frac{f}{\rho^2}\right)
   + \frac{m\omega_{\perp}^2}{2}\rho^2f+ gnf^3 = \mu_{\rm cell}(n(R_j,z)) f.
  \label{gps}
\eeq
This equation describes the vortex structure for all values of parameters,
provided that $N_v$ is large.  Equations (\ref{eprime}) and (\ref{gps})
generalize the result of Ref.  \cite{FG} through inclusion of the $(\nabla
\sqrt n\,)^2/2m$ and $m\omega_{\perp}^2\rho^2f^2/2$ terms.  The $(\nabla \sqrt
n)^2$ term allows us to go beyond Thomas-Fermi, when this energy dominates the
interaction term.  In the limit $\Omega \gg \omega_{\perp}^2/2gn$, appropriate
to the regime described in Ref.~\cite{FG}, the $\omega_{\perp}^2$ term in
(\ref{gps}) can be neglected.

It is useful to define the averages over the unit cell,
\beq
  a= \frac12 \ell^2\left\langle \left(
  \frac{\partial f}{\partial\rho}\right)^2
  + \frac{f^2}{\rho^2}\right\rangle,
\eeq
\beq
  a_h = \frac{1}{2\ell^2}\langle \rho^2(f^2-1)\rangle
\eeq
and
\beq
  b =  \langle f^4 \rangle;
\eeq
these quantities are dependent on the density within the cell.  Then quite
generally,
\beq
  E_j =  \int_j d^3r n\left\{\Omega \left(a-\frac34\right) +
                 \frac{\omega_{\perp}^2}{\Omega}a_h
                      + \frac{gn}{2}b \right\}.
\eeq

    For slow rotation, the core structure is basically that of a
single vortex \cite{lev}, and is reasonably well approximated by \cite{sandy}
\beq f\sim \frac{\rho}{(2\xi_0^2+ \rho^2)^{1/2}}, \label{sandy} \eeq where
$\xi_0 = \hbar/\sqrt{2mgn}$ is the Gross-Pitaevskii healing length.  The
corresponding density within this approximation to $f$ is shown as curve $a$
in Fig. 1 for the particular value, $\xi_0= 0.1\ell $.

\begin{figure}
\begin{center}
\epsfig{file=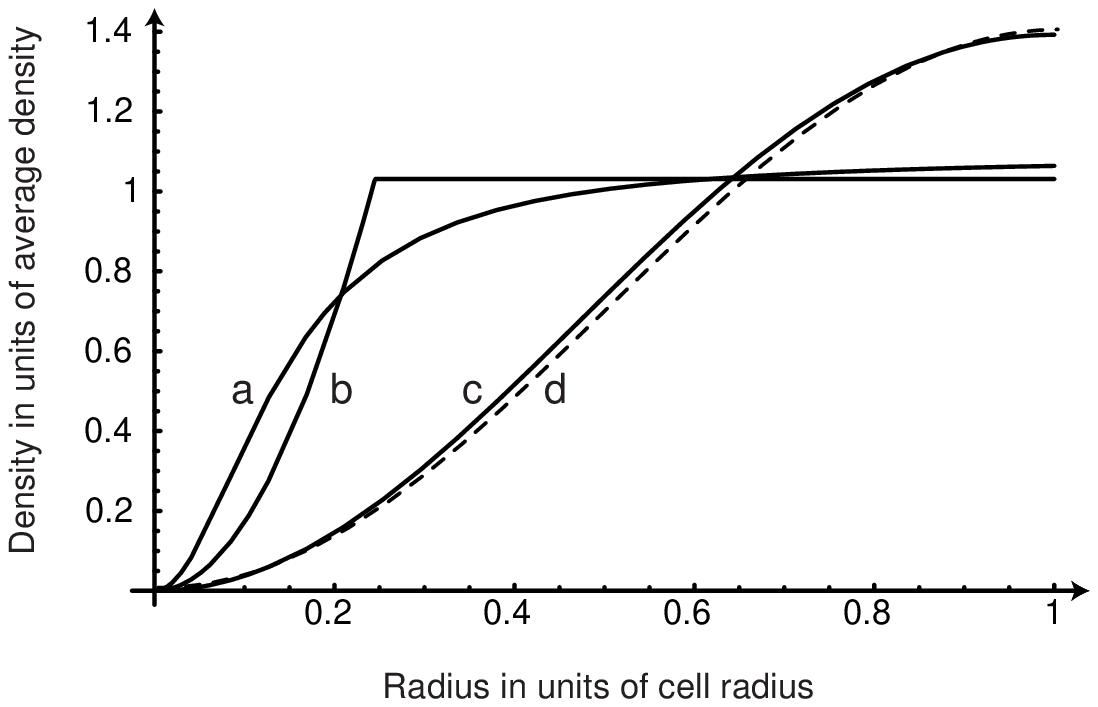,height=6.5cm}
\vspace{12pt}
\begin{caption}
{Density within a vortex core in units of the average density in
the cell, as a function of the transverse radius in units of the
core radius $\ell$:  a) the single vortex form, (\ref{sandy}); b) the
linear core approximation, (\ref{fb}); and c) the quantum Hall structure,
(\ref{qhf}), d) the free particle Bessel function $J_1$ (dashed).  Curves $a$
and
$b$ are calculated for $\xi_0=0.1\ell$.
}
\end{caption}
\end{center}
\label{FIG1}
\end{figure}

    Reference~\cite{FG} used a simple linear approximation for $f$ for all
rotation speeds, in which $f$ rises linearly to the effective core radius
$\xi$ and then becomes constant to the edge of the cell,
\renewcommand{\arraystretch}{1.5}
\beq
 f(\rho) = \frac{1}{(1-\xi^2/2\ell^2)^{1/2}}\times\left\{\begin{array}{l}
   \rho/\xi,
 \qquad 0 \le \rho \le \xi, \\
  1, \qquad  \xi \le \rho \le \ell.
\end{array}\right.
\label{fb}
\renewcommand{\arraystretch}{1.0}
\eeq
The corresponding density is shown as curve $b$ in Fig. 1, for the value
$\xi = \sqrt 6 \xi_0$ with $\xi_0 = 0.1\ell$.  In general one can solve
Eq.~(\ref{gps}) numerically for $s$, although we shall not do this here.

    With the linear approximation (\ref{fb}), the individual vortex energy,
$E_j$, becomes
\beq
   E_j = \int_j d^3r n\left(
    \Omega  a(\zeta)+ \frac{\omega_{\perp}^2}{\Omega}
  a_h(\zeta)+\frac{gn}{2}b(\zeta)
  -\frac{3\Omega}{4}  \right),
 \label{ej}
\eeq
where
\beq
   a(\zeta) =\frac{1-\frac12\ln\zeta}{1-\zeta/2},\qquad
   a_h(\zeta) =\frac{3\zeta-2\zeta^2}{12(2-\zeta)},
\eeq
and $\zeta = \xi^2/\ell^2$ is the fractional area occupied by the vortex
core.  The fluctuations in the density within a cell renormalize the (long
wavelength) coupling constant \cite{FG} by a factor $b = \langle n^2\rangle/
\langle n\rangle^2 >1$, given, for the ansatz (\ref{fb}), by
\beq
   b(\zeta) =\frac{1-2\zeta/3}{(1-\zeta/2)^2}.
\eeq

    The relative area occupied by the core at position $(r_\perp,z)$ is found
by minimizing the integrand of (\ref{ej}) at the density $n(r_\perp,z)$:
\beq
 \frac{\partial}{\partial \zeta}\left(a(\zeta)+ \frac{\omega_{\perp}^2}{\Omega^2}
  a_h(\zeta)\right)
  +\frac{gn}{2\Omega} \frac{\partial}{\partial \zeta}b(\zeta)=0.
 \label{min}
\eeq
In the Thomas-Fermi regime, the sound velocity, $s$, in the center of the
trap is given by
\beq
  ms^2 = gbn(0) =
\frac{\omega_{\perp}}{2}\left[\frac{15Nba_s}{d_\perp}\frac{\omega_z}{\omega_{\perp}}
\left(1-\frac{\Omega^2}{\omega_{\perp}^2}\right)
\right]^{2/5},
 \label{trapdensity}
\eeq
where $d_{\perp} =1/(m\omega_{\perp})^{1/2}$ is the oscillator length for
transverse motion.  We show, in Fig. 2, the corresponding prediction for
$\zeta$ at the center of the trap as a function of rotational velocity for
$^{87}$Rb, taking the representative values, $N=2.5\times10^6$,
$\omega_{\perp}/2\pi = 8.3$ Hz, and $\omega_z/2\pi = 5.2$ Hz.

\begin{figure}
\begin{center}
\epsfig{file=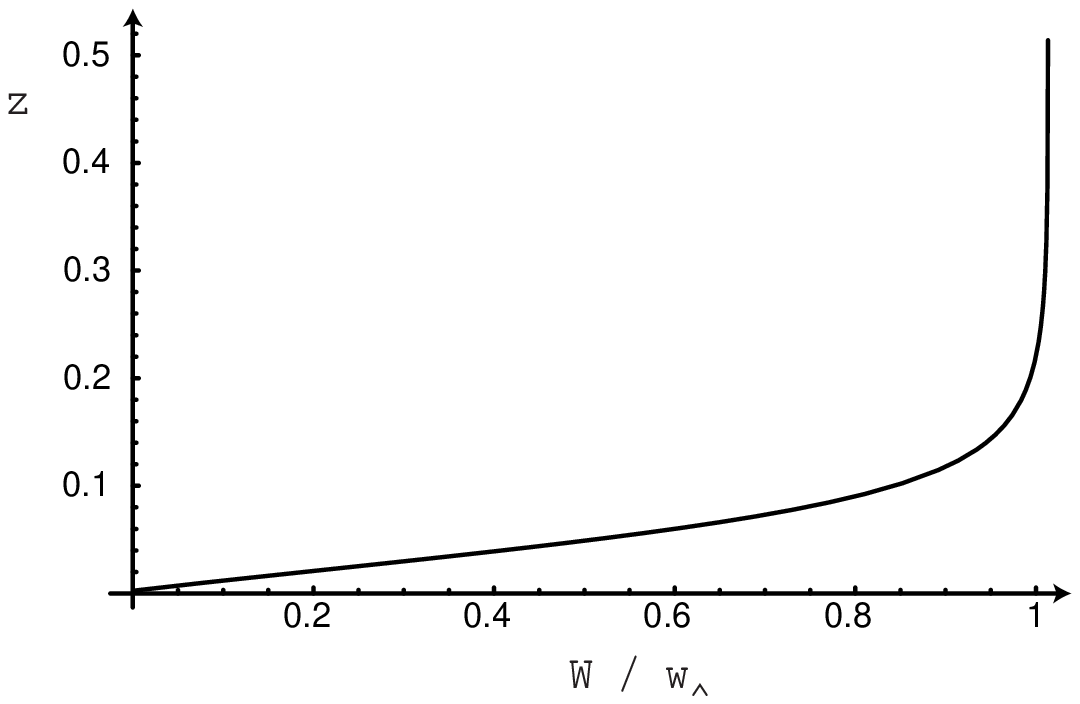,height=6.5cm}
\begin{caption}
{Variation of the core size with rotational velocity, in the linear
   approximation to the core structure.}
\end{caption}
\end{center}
\label{FIG2}
\end{figure}

    As we see in Fig.~2, the core structure changes rapidly as $\Omega$
approaches the transverse trap frequency $\omega_{\perp}$.  In order to study
rotational velocities comparable to the transverse trapping frequency it is
useful to spread out the horizontal scale by measuring rotational rates in
terms of the {\it rotational rapidity}, $y$, defined by \cite{rapidity}:
\beq
  \frac{\Omega}{\omega_{\perp}} \equiv \tanh y,
  \label{y}
\eeq
or
\beq
  y = \frac12 \ln \frac{\omega_{\perp}+\Omega}{\omega_{\perp} - \Omega}.
\eeq
The rapidity variable essentially counts the number of 9's in the fraction
$\Omega/\omega_{\perp}$ as the fraction approaches unity (just as metal
dealers describe the purity of metals).  For example, the currently achieved
\cite{jila2} $\Omega/\omega_{\perp} = 0.995$ corresponds to a rapidity of
3.00, while $\Omega/\omega_{\perp} = 0.999$ corresponds to $y=3.45$, and
$\Omega/\omega_{\perp} = 0.9999$ to $y=4.61$.  In Fig. 3 we show the variation
of the core size in Fig. 2, now as a function of rapidity.

\begin{figure}
\begin{center}
\epsfig{file=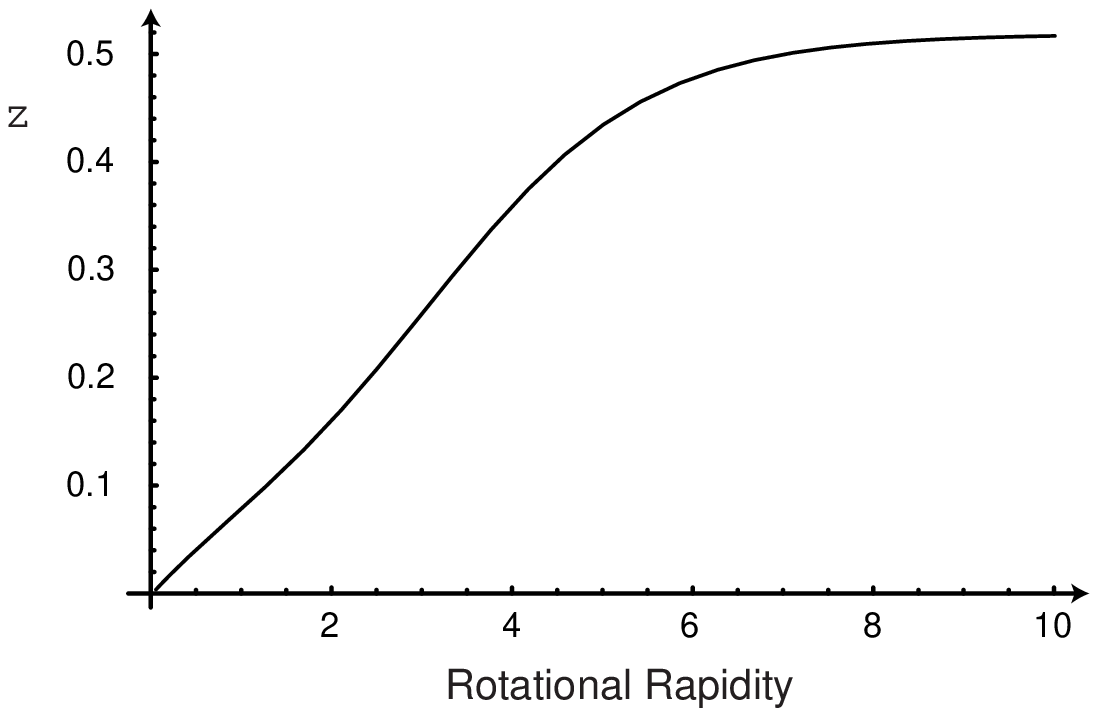,height=6.5cm}
\begin{caption}
{Variation of the core size shown in Fig. 2, as a function of
rotational rapidity, $y$.}
\end{caption}
\end{center}
\label{FIG3}
\end{figure}

    In the mean field quantum Hall regime, where $\Omega$ approaches
$\omega_{\perp}$, the cloud expands to the point where the $gn$ term becomes a small
perturbation on the structure within a cell; to lowest order $f$ assumes the
particularly simple oscillator p-state structure,
\beq
  f= C\frac\rho\ell e^{-\rho^2/2\ell^2},
 \label{qhf}
\eeq
plus small terms, where $C=(1-2/e)^{-1/2}$.  The p-wave solution for a
particle in a trap with frequency $\Omega$ which satisfies the usual boundary
condition $f\rightarrow 0$ for $\rho \rightarrow \infty$ kindly has zero slope
precisely at $\rho = \ell$.  With this form of $f$,
\beq
  E_j = \int_j d^3 r n\left(\Omega
   +\frac{m}{2}(\omega_{\perp}^2-\Omega^2)\rho^2 (f^2-1) + \frac g2 n f^4\right),
 \label{ejqh}
\eeq
so that
\beq
 E' = \int d^3r \left(\frac{(\nabla\sqrt{n})^2}{2m}
       + \frac{m}{2}n\omega_z^2 z^2 +\frac{gn^2b}{2}\right)
     +  N\Omega + \frac12 (\omega_{\perp}^2-\Omega^2)I;
\label{eprime5}
\eeq
this energy becomes the exact quantum Hall result in the limit, $\Omega
\to \omega_\perp$.  The average of $\rho^2/\ell^2$ is given by
\beq
   \frac{\int d^2\rho (\rho^2/\ell^2) f^2}{\int d^2 \rho}
   \equiv b' = \frac{2e-5}{e-2} = 0.608;
\eeq
in the linear core approximation one finds instead, 0.614.
The renormalization of the coupling constant by fluctuations in the
density within a cell, the factor $b = \langle n^2\rangle/ \langle
n\rangle^2$, is given in the quantum Hall limit by \cite{SHM,qhmodes},
\beq
 b= \frac{\int d^2 \rho f^4}{\int d^2 \rho}
   =\frac14\frac{e^2-5}{(e-2)^2} = 1.158;
\eeq
by comparison, the linear core approximation yields 1.192.  Note that the
total moment of inertia, $I$, equals $\bar I + N(b'-1/2)/\Omega$.

\section{Global Structure of the System}

    We turn now to determining the global structure of the cloud, by
minimizing
\beq
 E'  = \int d^3r \left(\frac{(\nabla\sqrt{n})^2}{2m}
       + \frac{m}{2}n\omega_z^2 z^2 \right)
   +\frac 12( \omega_{\perp}^2-\Omega^2 )\bar I +\sum_j E_j,
\label{eprime3}
\eeq
at fixed particle number.  From Eq.~(\ref{eprime3}) we derive the
effective Gross-Pitaevskii equation for the smoothed density:
\beq
  -\left\{\frac{\nabla^2}{2m}   + \frac m2 \left(\omega_z^2 z^2 +
   (\omega_{\perp}^2-\Omega^2 )r_\perp^2\right) +\mu_{\rm cell}(n)\right\}\sqrt n =
          \mu\sqrt n,
\eeq
where $\mu_{\rm cell} = \delta E_j/\delta n$, defined by Eq.~(\ref{gps}),
contains the explicit interaction energy term $gnb$.  In the quantum Hall
limit, Eq.~(\ref{ejqh}) implies that
\beq
   \mu_{\rm cell} = 2\Omega+ bgn+ b' \frac{\omega_{\perp}^2-\Omega^2}{2\Omega}.
\eeq

    The structure in the axial direction will, for $\Omega$ sufficiently close
to $\omega_{\perp}$, always become Gaussian.  The criterion for the axial structure to
be Gaussian is that $gn$ be small compared with the axial oscillator
frequency, $\omega_z$.  Since the system density falls indefinitely with
increasing $\Omega$, this condition will eventually be satisfied.  From
Eq.~(\ref{trapdensity}), the criterion is
\beq
    1-\frac{\Omega^2}{\omega_{\perp}^2} \ll
    \left(\frac{\omega_zb}{\omega_{\perp}}\right)^{3/2}
   \frac{2^{5/2}d_\perp}{15Na_s}
\eeq
or in terms of rapidity,
\beq
   y \gg \frac12 \ln\left[\left(\frac{\omega_{\perp}}{\omega_z}\right)^{3/2}
    \frac{15Na_s}{2^{1/2}d_\perp}\right].
\eeq

    For very weak interaction the transverse structure is Gaussian in the
quantum Hall limit \cite{Ho}:  \beq n(r_\perp,z) = {\pi \sigma(z)^2}
e^{-r^2/\sigma(z)^2} {\cal N}(z), \eeq where ${\cal N}(z)$ is the number of
particles per unit length in the axial direction.  As we shall see below, such
a Gaussian describes the system only for $Na_s \ll d_z$, where $d_z =
1/(m\omega_z)^{1/2}$ is the axial oscillator length.  For this Gaussian,
\beq
 E'=\int dz \left\{\frac{1}{2m}\left(\frac{d({\cal N}(z))^{1/2}}{dz}\right)^2
         +{\cal N}(z)\left(\frac{1}{2m\sigma(z)^2}
   +  \frac m2 \left(\omega_z^2 z^2 +
   (\omega_{\perp}^2-\Omega^2 )\sigma(z)^2\right)
   +\frac{bg}{4\sigma(z)^2}{\cal N}(z)\right)\right\},
\eeq
plus a constant times $N$.  Minimizing with respect to $\sigma(z)$ at
fixed ${\cal N}(z)$, we find,
\beq
     \sigma(z) = d\left(1+  2\pi b {\cal N}(z)a_s\right)^{1/4}
            \left(\frac{\omega_{\perp}^2}{\omega_{\perp}^2-\Omega^2}\right)^{1/4},
\eeq
in agreement with Ref.~\cite{Ho}.

    However, if the transverse structure of the non-rotating cloud is
Thomas-Fermi, it will remain Thomas-Fermi as the cloud is spun up, even to the
quantum Hall limit.  The criterion for the transverse structure to be Gaussian
is different than in the axial direction, since the effective transverse
oscillator frequency, $(\omega_{\perp}^2-\Omega^2)^{1/2}$, goes to zero.  The
criterion becomes instead that the interaction energy, $gn$, be small compared
with the transverse kinetic energy:  $gn \sim gN/ZR_\perp^2 \ll
1/2mR_\perp^2$, where $Z$ is the axial height and $R_\perp$ the transverse
radius.  This condition implies that $Na_s/Z$ be $\ll 1$.  Since the total
density per unit height, $N/Z$, increases with increasing $\Omega$ as the
system flattens out, the structure in the transverse direction can only be
Gaussian if the transverse structure in the non-rotating cloud is itself
Gaussian.  The maximum that $N/Z$ can become is $\sim N/d_z$, where $d_z$ is
the axial oscillator length.  For $Na_s/d_z \gg 1$ the structure in the radial
direction will be Thomas-Fermi at large $\Omega$, even if it is Gaussian at
small $\Omega$.

    Note that in the quantum Hall limit, even though the interaction energy
plays only a perturbative role in determining the structure within each cell
of the lattice, it is crucial in determining the global structure.  In
particular, it is responsible for inclusion of components from higher Landau
levels required to produce a Thomas-Fermi profile.

    The final axial-Gaussian, transverse-Thomas-Fermi structure at high
rotation has the form
\beq
    n(\vec r\,) =
e^{-z^2/d_z^2}\left(n(0)-\frac{m}{2gb}(\omega_{\perp}^2-\Omega^2)r_\perp^2\right).
\eeq
Using $\int d^3 r n = N$, we find
\beq
  N = \frac{\pi^{3/2}}{2}d_z R_\perp^2 n(0),
\eeq
where the transverse size, $R_\perp$, is given by the point where $n(\vec
r\,)$
falls to zero,
\beq
  R_\perp = \left( \frac{2gbn(0)}{m(\omega_{\perp}^2-\Omega^2)}\right)^{1/2} =
\left(8\pi b a_s d^2 n(0)\right)^{1/2}
\left(\frac{\omega^2}{\omega_{\perp}^2-\Omega^2}\right)^{1/2}.
\eeq
In terms of the total number, $N$,
\beq
  \frac{R_\perp}{d} = \frac{2}{\pi^{1/8}}\left(Nb
   \frac{a_s}{d_z} \,\frac{\omega_{\perp}^2}{\omega^2-\Omega^2} \right)^{1/4},
\eeq
and
\beq
   n(0)=  \frac{1}{2\pi^{5/4}} \left(\frac{N}{b d^4d_z a_s}
       \,\frac{\omega^2-\Omega^2}{\omega_{\perp}^2} \right)^{1/2}.
\eeq

\section{Measuring the Core Size}

    Several quantitative measures can be used to compare predicted core sizes
with experiment, and with theory in the quantum Hall regime.  The first is
simply to compare the slopes of $f$ at the origin.  The slope of the order
parameter in the linear approximation, Eq.~(\ref{fb}), is
$1/\ell(\zeta(1-\zeta/2))^{1/2}$, which approaches $1.62/\ell$ as $\Omega \to
\omega_{\perp}$.  On the other hand the quantum Hall wave function, (\ref{qhf}), has
slope $1/\ell(1-2/e)^{1/2} = 1.95/\ell$.  The second is to measure the mean
square radius, $r_c^2$, of the density deficit in the core, defined by,
\beq
   r_c^2 = \frac{\int_j d^2 \rho \left[f(\ell)^2 - f(\rho)^2\right]\rho^2}
   {\int_j d^2 \rho \left[f(\ell)^2 - f(\rho)^2\right]}.
\eeq
For the quantum Hall wave function, $r_c^2/\ell^2 = ((11/2)-2e)/(3-e)
\simeq 0.225$.  To compare with the result from the linear approximation to
$f$, we note that $r_c^2/\ell^2 = \zeta/3$, while as $\Omega$ approaches
$\omega_{\perp}$, the value of $\zeta$ is found from the minimum of
$a(\zeta)+a_h(\zeta)$, which is at $\zeta \simeq 0.519$; thus in this limit,
$r_c^2/\ell^2 \simeq 0.173$.  Note that although the initial slope of the
quantum Hall wave function is larger than that in the linear approximation,
the mean square radius of the depression is also larger, since the depression
in the quantum Hall wave function extends over the entire cell.  Both measures
of the core size in the linear approximation are in reasonable agreement with
the exact quantum Hall result, given the simplicity of the approximation.

    Experimentally, core properties are investigated after the rotating cloud
has expanded.  In the JILA experiments, the atoms are transferred to a state
in which the magnetic forces tend to drive the cloud apart.  It is therefore
necessary to investigate how the vortex-core structure is affected by
the transfer to the new state and the subsequent expansion of
the cloud.  Under expansion, the density drops, eventually reaching the
point where the interaction energy no longer plays a role in determining the
structure within the individual cells.  The centrifugal force plays no role
within a cell.  If the potential is adiabatically turned off, allowing the
system to expand slowly to the point where the interaction within a cell is
small compared with the bending energy of the order parameter within a cell,
or $1/2m\ell^2 = N_v/2mR_\perp^2 \gg gN/R_\perp^2Z$. which is the case
when the axial height expands to the point where $Z \gg 8\pi a_s N/N_v$,
then the structure within the individual cells, given by Eq.
(\ref{gps}), is the Bessel function, $C_1 J_1(x_0 r/\ell')$, where $x_0 =
1.84$ is the location of the first maximum of the $J_1(x)$, $C_1 = 2.05$ and
$\ell'$ is the cell size in the expanded cloud.  In fact, the Bessel
function solution is always within 0.015 of the quantum Hall solution, and
the two solutions would be effectively indistinguishable in practice (see
Fig. 1).  The slope at the origin of the Bessel function is 1.88, compared
with 1.85 for the quantum Hall solution, while the mean square radius,
$r_c^2/\ell^2$, of the depression is 0.231, compared with 0.225 for the
quantum Hall solution.

    One can distinguish two stages in the evolution of the cloud during
release and the subsequent expansion.  The first is the period when the atoms
are transferred to an untrapped state, and the second is expansion of the
cloud in a modified trapping potential.  The transfer of atoms occurs on a
time scale short compared with dynamical times for the particles.  Therefore
the sudden approximation should be good, and changes in the coordinate-space
wave function during the transfer should be negligible.  This implies that
both the global structure of the cloud and the structure of an individual cell
of the vortex lattice are unchanged.  After transfer of atoms to the new
state, the structure within a cell will not correspond to the equilibrium
configuration for the particular rotation rate because of the change in the
trapping potential, which is determined by the instantaneous value of
$\omega_{\perp}$.  The calculations described in the previous paragraph
demonstrate that when interaction effects are small, the structure of the
condensate wave function within a single cell depends only weakly on
$\omega_{\perp}$.  Therefore, after transfer, the wave function within a cell
will be the lowest state for the new value of $\omega_{\perp}$, apart from
corrections of order one per cent.  Likewise, for rotation rates so small that
interaction effects dominate, we expect a similar conclusion to hold because
the oscillator potential plays little role in determining the structure of an
individual vortex.

    We now consider the degree to which the vortex cores adjust adiabatically
in the expansion.  To do this, we compare the time scale, $\tau_{\rm cell}$,
for response of the structure of a cell of a vortex lattice with the expansion
time scale, $\tau_{\rm exp}$.  When the vortex core radius is small compared
with the cell radius, the time for adjustments of the core is of order the
core radius, $\sim (mgn)^{-1/2}$, divided by the sound speed $s$, or
$\tau_{\rm cell} \sim \hbar/gn$.  When the core radius becomes comparable to
the cell radius, i.e., $\hbar \Omega \simge gn$, the inverse response time
becomes of order the kinetic energy associated with a particle confined within
a volume of radius $\ell$, divided by $\hbar$, or $\tau_{\rm cell} \sim
m\ell^2/\hbar = 1/\Omega$.  Thus $1/\tau_{\rm cell}$ is always the larger of
$gn/\hbar$ and $\Omega$.  These estimates should apply at all stages in the
evolution, provided that $n$ and $\Omega$ are the instantaneous values of
these quantities.  We note that if the expansion is purely two-dimensional, a
good approximation for the recent experiments \cite{jila2}, the density and
$\Omega$ both scale as $1/R_\perp^2$; therefore, the ratio $gn/\hbar \Omega$
remains constant, and the core expansion rate always remains $gn/\hbar$ or
$\Omega$.

    There are similarly two regimes for the expansion.  At low rotation rates,
when the interaction energy per particle, $gn$, is large compared with $\hbar
\omega_\perp$, the expansion velocity is determined by the interaction energy
of the cloud, and is typically of order the sound velocity, $s_0$, in the
cloud before release (the subscript $0$ denotes quantities just prior to
release).  On the other hand, when the typical initial orbital velocity,
$\Omega_0 R_{\perp 0}$, exceeds the sound velocity, $s_0$, the dominant
contribution to the expansion velocity after switching off the trap potential
is the orbital motion, and therefore the expansion velocity is of order
$\Omega_0 R_{\perp 0}$.  The typical expansion rate, $1/\tau_{\rm exp}$, is
thus always the larger of $s_0/R_\perp$ and $\Omega_0 {R_{\perp 0}}/R_\perp$.

    Now let us compare time scales.  For low rotation velocities, $\Omega_0 \,
\la \, s_0/R_{\perp 0}$, we have
\beq
  \frac{\tau_{\rm cell}}{\tau_{\rm exp}}
  \sim \frac{1}{ms_0 R_{\perp 0}}\frac{n_0 R_{\perp 0}}{nR_{\perp}}
  \sim \frac{1}{ms_0 R_{\perp 0}}\frac{R_\perp}{R_{\perp 0}},
\eeq
where the latter estimate holds for two dimensional expansion.  This ratio
is initially smaller than unity, implying that the cell initially
adiabatically adjusts during the expansion, but if the cloud expands to a
radius $\simge R_{\perp 0}^2/\xi_0$, where $\xi_0$ is the
Gross-Pitaevskii healing length, the condition for adiabaticity
will be violated.  For intermediate rotation rates, $s_0/R_{{\perp} 0}\, \la
\Omega_0 \la \, gn_0/\hbar$, the ratio of times is given by
\beq
 \frac{\tau_{\rm cell}}{\tau_{\rm exp}} \sim \frac{\hbar
 \Omega_0}{gn}\frac{ R_{\perp 0}}{R_\perp}.
\eeq
This ratio starts at a value less than unity but increases $\propto
R_\perp/R_{\perp 0}$ as the cloud expands.  For the final case of fast
rotation, $\Omega_0\ga gn_0/\hbar$, the ratio is
\beq
\frac{\tau_{\rm cell}}{\tau_{\rm exp}} \sim
 \frac{\Omega_0 R_{\perp 0}}{\Omega R_\perp} \sim \frac{R_\perp}{R_{\perp 0}}.
\eeq
In this case the adiabatic assumption is marginally satisfied initially,
and is violated during the subsequent expansion.  We conclude that one may
draw no general conclusions about the development of vortex core structure
during expansion on the basis of arguments about time scales; more detailed
studies are needed.

    It is interesting to note that states made up only of components in the
lowest Landau level expand homologously when the effects of interaction are
neglected \cite{read}.  In this case the structure of a single cell remains
invariant, with only changes in scale, independent of the transverse length
entering the wave function [cf.  Eq.~(\ref{qhf})].  Even though the condition
for adiabaticity is violated, the structure of the single cell is precisely
what would be predicted assuming adiabatic behavior.

\section{Conclusion}

    In this paper we have developed a unified framework for describing the
structure of rotating Bose-Einstein condensates containing a large number of
vortices.  We have derived a Gross-Pitaevskii equation which describes the
structure of individual vortices and have demonstrated how the mean field
quantum Hall state emerges as a simple continuation of the structure for small
rotation rates.  We find that the global density profile of the rotating
clouds in the transverse direction is generally of the Thomas-Fermi form,
rather than the Gaussian that emerges if only the lowest Landau level is
occupied.

    A number of open problems remain for future work.  Throughout, we have
assumed that the Gross-Pitaevskii approach may be used, and have neglected
effects of excited states.  One such effect is the zero-point motion of
collective modes \cite{tkmodes,qhmodes}, which broadens the density profile of
individual vortices and makes the lowest density non-zero.  The density is the
center of the vortex can also become non-zero via anomolous modes of
excitation of the condensate \cite{fetterreview} that in the linear
approximation have a negative excitation energy.  Even at zero temperature,
such modes will have a non-zero equilibrium population such that the energy of
an anomalous modes, including the effects of self-interaction, is just equal
to zero.  Explicit calculations are given in Ref.~ \cite{salomaa}.  A third
effect is the thermal population of excited states, which likewise will lead
to a non-zero density at the center of the vortex.  All of these effects must
be taken into account in a detailed comparison of experiment with theory.
Further problems include the quantitative delineation of the effect of
expansion on the vortex core structure, and inclusion of effects of the
lattice beyond the Wigner-Seitz approximation, such as the rigidity to shear
motion, which manifests itself, e.g., in Tkachenko modes
\cite{Tkachenko,jilatk,tkmodes}.

    We thank James Anglin, Jason Ho, and Volker Schweikhard for many good
discussions at the Aspen Center for Physics, to which we are grateful for
giving us the opportunity to carry out this research.  This work was supported
in part by NSF Grant PHY00-98353.


\begin{references}
\bibitem{corrections} When the number of vortices is finite, there is a
net flow of fluid past the vortices.  Thus $m\Omega/\pi$ will differ slightly
from the vortex density, $N_v/\pi R_\perp^2$.  The relative flow past the
vortex cores produces a net Magnus force on individual vortices, of relative
order $1-mR_\perp^2\Omega /N_v$, leading, e.g., to small distortions of the
lattice from triangular.
\bibitem{rapidity} We make this definition in direct analogy with the
common concept of rapidity in high energy physics, $v/c = \tanh y$, where $v$
is the particle velocity and $c$ is the speed of light, introduced to
distinguish speeds of ultrarelativistic particles.
\end{references}
\end{document}